\documentclass[twocolumn]{svjour3}
\usepackage{graphicx,amsmath,amsfonts}
\usepackage{textcomp}
\usepackage{units}
\usepackage{epsfig}
\usepackage{color}

\newcommand{\Yb}{$^{171}$Yb$^+$ }

\newcommand{\nm}[1]{\unit[#1]{nm}}
\newcommand{\mum}[1]{\unit[#1]{\textmu m}} 
\newcommand{\mm}[1]{\unit[#1]{mm}}

\newcommand{\MHz}[1]{\unit[#1]{MHz}}
\newcommand{\kHz}[1]{\unit[#1]{kHz}}
\newcommand{\ka}[1]{\unit[#1]{cm$^{-1}$}}

\newcommand{\ybodd}{$^{171}$Yb }
\newcommand{\ybeven}{$^{172}$Yb }

\newcommand{\Shalf}{S$_{1/2}$ }
\newcommand{\Szero}{$^1$S$_0$ }
\newcommand{\Phalf}{P$_{1/2}$ }
\newcommand{\Pone}{$^1$P$_1$ }

\begin{document}
\graphicspath{{Images/}}
\title{Resonance enhanced isotope-selective photoionization of YbI for ion trap loading}
\author{M. Johanning \and A. Braun \and D. Eiteneuer \and Chr. Paape \and  Chr. Balzer \and W. Neuhauser
\and Chr. Wunderlich
}
\institute{M. Johanning \and A. Braun \and D. Eiteneuer \and  Chr. Balzer \and Chr. Wunderlich \at Fachbereich Physik, Universit\"at Siegen, 57068 Siegen, \\\email{wunderlich@physik.uni-siegen.de}
Germany \and Chr. Paape \and  W. Neuhauser \at Institut f\"ur Laser-Physik, Universit\"at Hamburg,
Luruper Chaussee 149, 22761 Hamburg, Germany}
\date{Received: date / Revised version: date}
\maketitle

\begin{abstract}
Neutral ytterbium (YbI) and singly ionized ytterbium (YbII) is
widely used in experiments in quantum optics, metrology and quantum
information science. We report on the investigation of isotope
selective two-photon-ionization of YbI that allows for efficient
loading of ion traps with YbII. Results are presented on two-colour
(\nm{399} and \nm{369}) and single-colour (\nm{399}) photoionization and
their efficiency is compared to electron impact ionization. Nearly
deterministic loading of a desired number of YbII ions into a linear
Paul trap is demonstrated.
\end{abstract}

\sloppy
\section{Introduction}
Atomic and singly ionized ytterbium is the element of choice in
numerous experiments in atomic physics and quantum optics. Trapping
and laser cooling of \Yb in Paul traps was initially motivated by
its potential use as a frequency standard in the microwave or
optical regime
\cite{Strumia1978,Blatt1982,Werth1986,Blatt1989,Klein1990,Fisk1993,Seidel1995,Gill2003,Schneider2005}.
In addition, this element is now employed by an increasing number of
research groups for implementing experiments related to quantum
information science (e.g.,
\cite{Huesmann1999,Hannemann2002,Wunderlich2003,Balzer2006,Kielpinski2006,Ozeri2007,Maunz2007,ChuangVuletic2007}).
Neutral Yb is used, for instance, in Bose-Einstein-condensation
\cite{Takasu2003}, photoassociation \cite{Takasu2004}, or for
experiments aiming at the measurement of parity violation in atoms
\cite{Natarajan2005}.

An important step in trapping ions is the ionization process. For
almost 25 years since the first successful trapping and laser
cooling of atomic ions \cite{Neuhauser1978,Wineland1978} electron
impact ionization was the process of choice. In recent years
photoionization has begun to establish itself as an attractive
ionization process for different elements, as it offers distinct
advantages over electron impact ionization
\cite{Mortensen2004,Kjaergaard2000,Gulde2001,Lucas2004,Brownutt2007,Deslauriers2006,Balzer2006}. First, it is
highly efficient in comparison with electron impact ionization,
typically yielding loading rates that are several orders of
magnitude higher. This allows to reduce the neutral atom flux
considerably, thus avoiding contamination by deposition of atoms on
trap electrodes or other, isolating trap components. Second, the
electron bombardment introduces excess charge into the trap volume
and on trap components that disturbs the trapping potential. This
effect, too, is avoided when using photoionization. And third, by
employing a resonantly enhanced process, photoionization can be made
isotope selective in contrast to indiscriminate trap loading with
the electron impact method. Thus, the need for isotope-enriched
samples, often a prerequisite for experiments with isotopically pure
ionic crystals, is eliminated.

In this article we present the investigation of photoionization of
atomic ytterbium, with particular attention to its use as a loading
process for ion traps. A further advantage of the scheme presented
here is the need for only one light source at a convenient
wavelength near \nm{399} that is reasonably close to the now
established industrial standard at \nm{405} for laser diodes making
access to inexpensive laser light sources more likely. Recent
progress in laser diode production processes, however, makes
outliers, which deliver the desired wavelength, less probable and
makes it necessary to use cooled laser diodes \cite{Kielpinski2006}.

In section~\ref{sec:ResFluor} we report on laser spectroscopy of the
\Szero - \Pone resonance near \nm{399} in YbI with the hyperfine
structure resolved. Resonance fluorescence spectra are recorded
employing a collimated thermal atomic beam emerging from an Yb-oven.
We present a simplified model of this Yb-oven in section~\ref{sec:ovenmodel} 
that predicts measured oven temperatures
satisfyingly and agrees with density and flux measurements presented
in section~\ref{sec:densityflux}. Resonance enhanced photoionization
of YbI using an additional light field near \nm{369} is then
reported in section~\ref{sec:twoColor} while in section~\ref{sec:oneColor} 
it is shown that the light field near \nm{399}
alone may ionize YbI in the presence of the rf field used for ion
trapping. Section \ref{sec:conclusion} concludes this article.

\section{Laser spectroscopy of neutral Yb}
\label{sec:ResFluor}

Photoionization of Yb by one photon would require radiation with
wavenumber larger than \ka{50443} \cite{Meggers1978} corresponding
to a wavelength below \nm{200} in the vacuum-uv range. Generating
and guiding radiation in this wavelength regime is rather
inconvenient. In addition, one-photon-ionization does not take
advantage of bound energy levels of Yb in order to make this process
isotope selective. Therefore it is useful to first excite Yb to a
bound excited state and subsequently ionize the excited atom.

\begin{figure}[htbp]
  \centering
  \epsfig{file=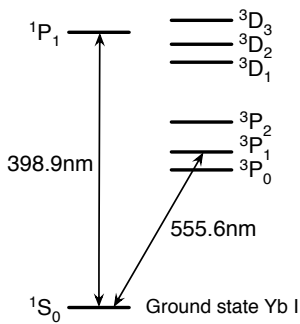, width=4cm}
  \caption{Partial energy level scheme of YbI. Two resonances convenient for laser
    excitation and subsequent photoionization are indicated by arrows.}
  \label{fig:levelScheme}
\end{figure}

Fig.~\ref{fig:levelScheme} depicts relevant low lying energy levels
in YbI. Starting from the \Szero ground state several energy levels
may serve as a first excitation level from which the actual
ionization process occurs. The inter-combination line near
\nm{555.6} has been extensively studied
\cite{Tkachev1996,Borisov1998} as starting
point for a photoionization process. Unfortunately, this wavelength
is presently not accessible using an inexpensive diode laser.
Therefore, we concentrate on the optical dipole transition from the
ground state to the \Pone level near \nm{398.9}. The \Pone state has
a lifetime of 5.464~ns \cite{Takasu2004} and the transition \Szero -
\Pone is saturated by a light intensity of 60 mW/cm$^2$
\cite{Maruyama2003}. Possible decay channels of the \Pone state are
not indicated in Fig.~\ref{fig:levelScheme}, as the ratio between
the rate for radiative decay into the ground state and the sum of
the rates for decay into all other states is larger than $10^7$
\cite{Migdalek1991}.
\begin{figure}[htbp]
  \centering
  \epsfig{file=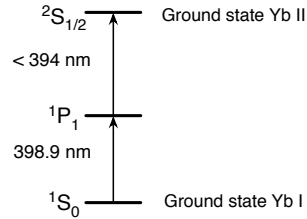, width=4cm}
  \caption{Two-photon-ionization of YbI via the \Pone state.}

  \label{fig:ionizationScheme}
\end{figure}

As indicated in Fig.~\ref{fig:ionizationScheme}, a second
light-field with wavelength below \nm{394} is required to reach the
continuum. This process has been proposed by Sankari et al.
\cite{Sankari1998}. Fortunately, this second light-field is easily
supplied by the cooling laser for the YbII ion near \nm{369} which
is already present at experiments working with trapped YbII. This
makes this scheme very practical indeed, as only one other diode
laser needs to be set up. This process is termed two-colour-ionization
and its experimental investigation is described in detail
in section~\ref{sec:twoColor}.

This contrasts with the scheme presented in section~\ref{sec:oneColor} 
that we label one-colour-ionization. As is shown
experimentally, a second light-field near \nm{369} is not necessary
to ionize YbI atoms in an ion trap, instead employing only the laser
near \nm{398.9} is sufficient.

In the remainder of this section we will be concerned with the
spectroscopy and isotope selective excitation of the \Szero - \Pone
transition in YbI.

Laser induced resonance fluorescence spectra of YbI are recorded
using a thermal atomic beam emitted from three different ovens
heated resistively. The ovens are molybdenum tubes with an inner
diameter of \mm{0.78} and a wall thickness of \mm{0.2}, spot-welded
to a tantalum wire.  The ground connection is cut from a tantalum foil
(thickness \mm{0.2}) with a size of approximately
\unit[$2\times4$]{mm$^2$}, spot-welded close to the opening of the
oven tube. Two ovens contain samples of YbI that are isotopically
enriched with \ybodd and \ybeven, respectively while the third oven
contains a sample with the natural abundance of isotopes of
ytterbium. Two different experimental set-ups are used for the
experiments reported in this article: The isotopically enriched
ovens are mounted in a vacuum chamber maintained at a pressure below
\unit[$10^{-10}$]{mbar} (when the oven is turned off) that also houses a
miniature Paul trap (compare section~\ref{sec:experiment}).  The
third oven is contained in a vacuum chamber at a typical pressure of
\unit[$5\cdot10^{-9}$]{mbar} together with a linear Paul trap
\cite{Balzer2006}.

For excitation of the \Szero - \Pone transition a tuneable External
Cavity Diode Laser is used with an emission bandwidth of about
\kHz{$2\pi \times 500$}. The wavelength of the laser radiation is
measured using a wave meter to a precision of \unit[0.05]{pm} using
a temperature stabilised HeNe laser as a reference corresponding to
an uncertainty of $\approx $\MHz{95} in absolute frequency. The
laser beam crosses the atomic beam at an angle close to 45$^\circ$
or 90$^\circ$.

Resonance fluorescence of YbI close to \nm{399} is collected,
imaged, and detected using the setups that are in use for photon
counting the resonance fluorescence close to \nm{369} originating
from the \Phalf - \Shalf transition in YbII. With the setup
containing the isotopically enriched YbI ovens light is collected
using an objective with numerical aperture of approximately 0.4 and
then imaged onto a photomultiplier tube (Burle C31034A2) operated in
photon-counting mode. The overall detection efficiency is about 1\%.
The setup with the oven containing ytterbium with natural abundance
of isotopes uses optical elements for light collection  with a
numerical aperture of approximately 0.5 and a photomultiplier
(Hamamatsu R5600P) with an overall calculated efficiency of 1.3\%
for the detection of light emitted by YbI and YbII.

\subsection{Isotope selective excitation of YbI}
\begin{figure}[htbp]
  \centering \epsfig{file=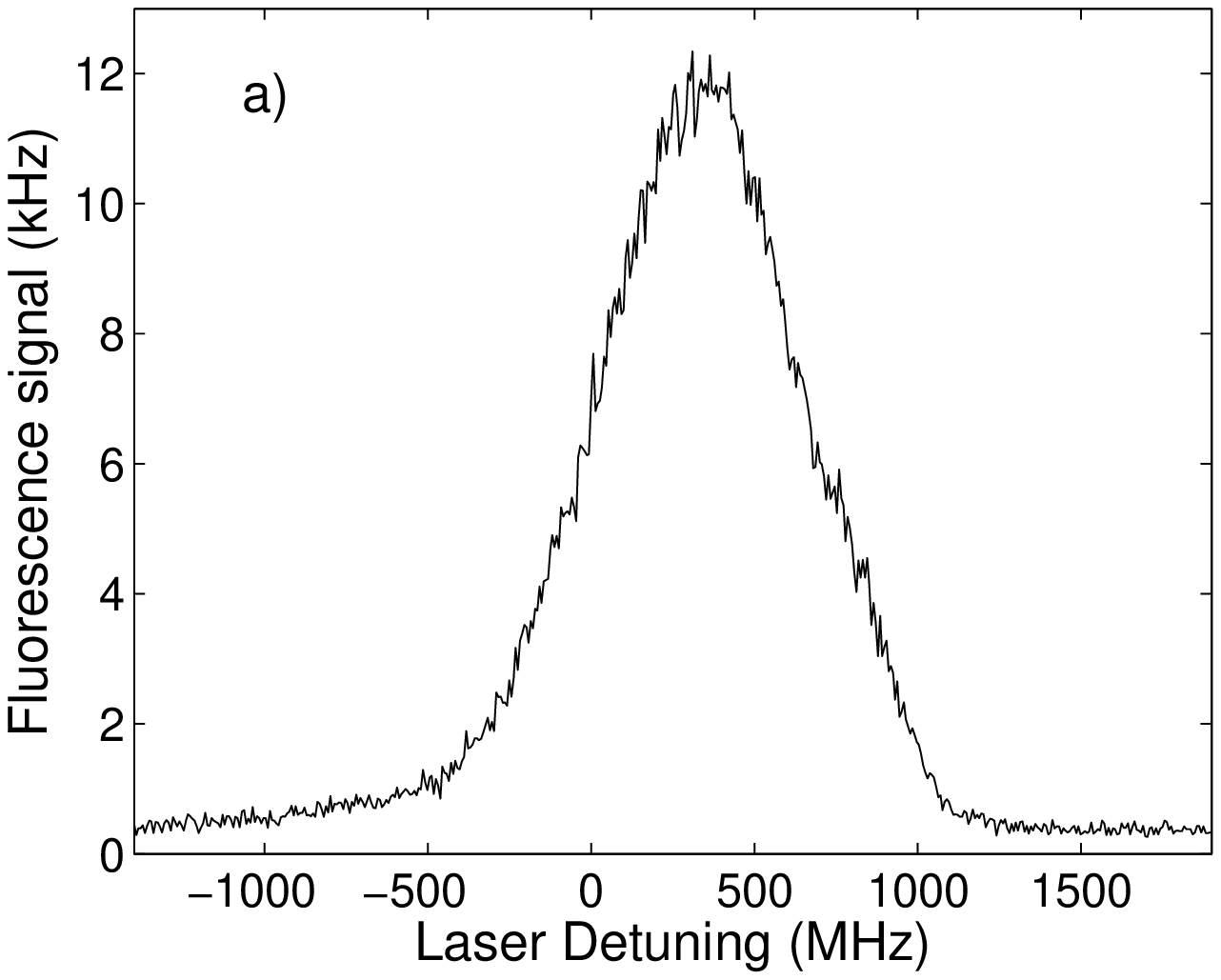, width=6cm}
  \centering \epsfig{file=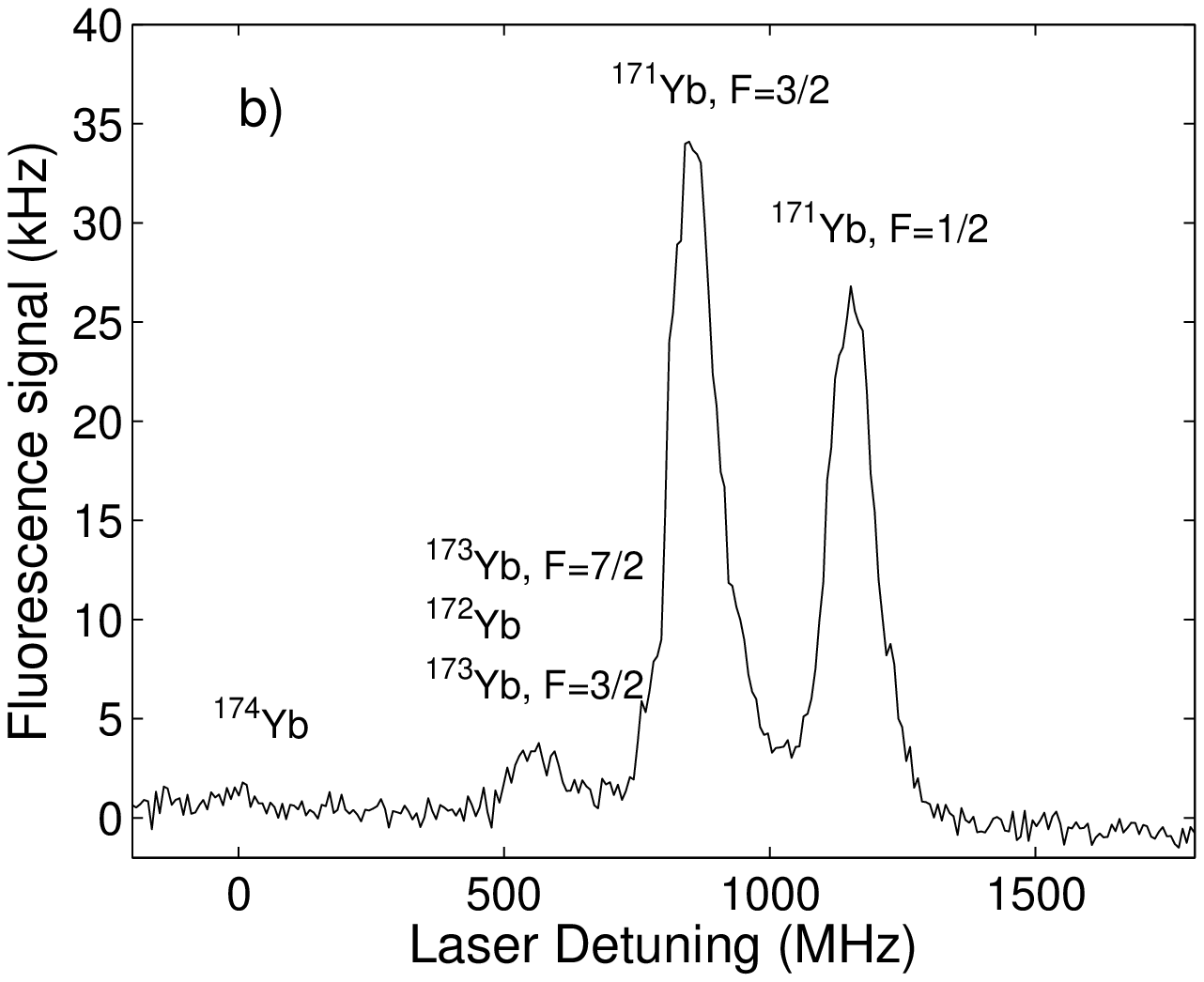, width=6cm}
  \centering \epsfig{file=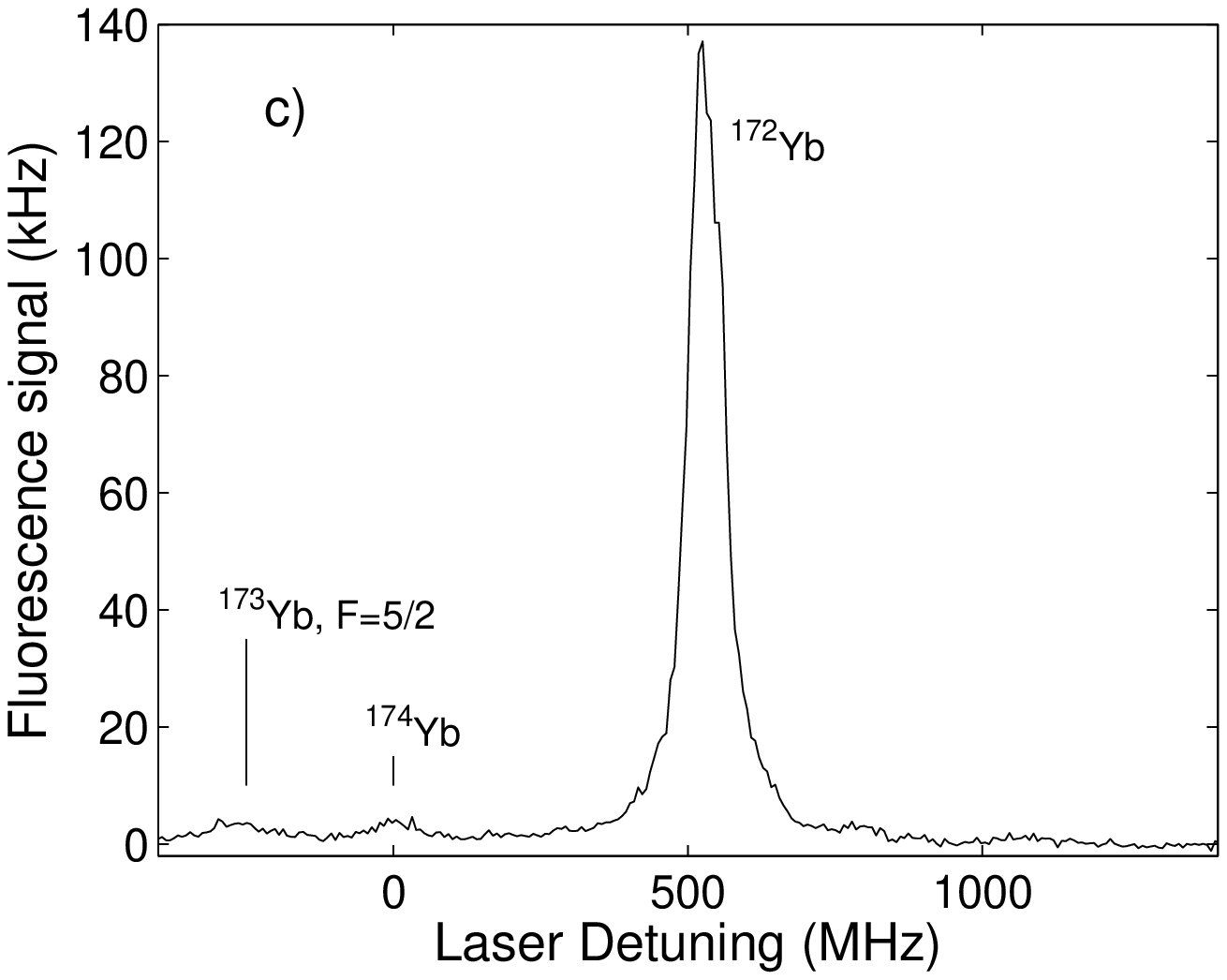, width=6cm}
  \caption{Laser induced resonance fluorescence spectra of the \Szero -
    \Pone transition in YbI.  An isotopically enriched sample of
    $^{171}$YbI (a, b) or $^{172}$YbI (c) was used. The detuning of the laser frequency is
    given relative to the resonance of $^{174}$Yb.
    a) $^{171}$YbI: The angle between laser beam and atomic
    beam is about 45 degrees.  The resonance is
    shifted and broadened by the Doppler effect.
    b)$^{171}$YbI: The laser beam crosses the
    atomic beam at right angle allowing for resolution of the hyperfine structure.
    The rest abundance of $^{172}$YbI can be seen as a small resonance peak to the left of
    the two prominent hyper-fine resonances of $^{171}$ YbI.
    c)$^{172}$YbI: Doppler-reduced spectrum.
    }
  \label{fig:excitationPureOven1}
\end{figure}

Figure~\ref{fig:excitationPureOven1} shows laser induced resonance
fluorescence spectra of the \Szero - \Pone transition in YbI. These
were measured employing isotopically enriched samples of \ybodd or
\ybeven. The abscissa indicates the frequency relative to the centre
of the resonance corresponding to $^{174}$ YbI that was found at a
vacuum wavelength near \nm{398.91}.

The excitation line profile in Fig.~\ref{fig:excitationPureOven1}a)
is Doppler-broadened and shifted while in
Figs.~\ref{fig:excitationPureOven1} b) and c) Doppler-broadening is
strongly reduced, with the measured linewidth of $\Gamma_{172}$ =
\MHz{$2\pi\cdot 71(1)$} and $\Gamma_{171}$ = \MHz{$2\pi\cdot 89(1)$} a
factor of approximately 3 higher than the natural linewidth of
\MHz{$2\pi\cdot 28$} due to saturation broadening and the Doppler-broadening due to the divergence
of the atomic beam. The hyperfine splitting
of the \Pone level of the \ybodd isotope is clearly resolved.
Small peaks to the left and right of the
respective main resonance in Figs.~\ref{fig:excitationPureOven1}
b) and c) are due to impurities from other isotopes.

\begin{figure}[htbp]
  \centering
  \epsfig{file=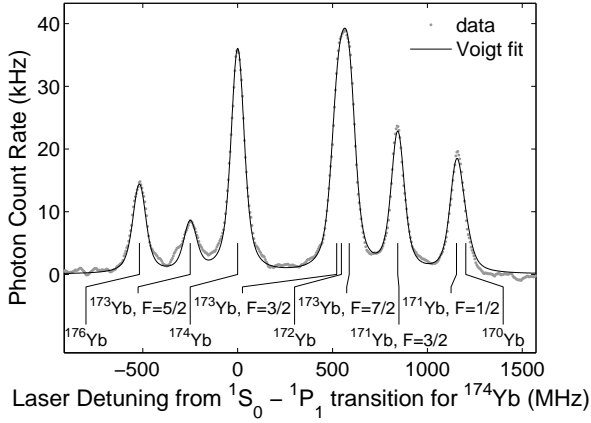, width=8cm}
  \caption{Laser induced resonance fluorescence spectra of the \Szero -
    \Pone transition in YbI with reduced Doppler-broadening from an
    oven with natural abundance isotope distribution. All hyperfine information displayed at the resonances concerns the upper levels of the transitions since the ground state exhibits no hyperfine splitting. Nonlinear regression was performed by assuming fixed spacing given by \cite{Das2005} and fitting a common horizontal and vertical displacement, a natural line width, and residual Doppler width, and individual peak heights.}
  \label{fig:excitationMixedOven}
\end{figure}

After having obtained sufficient spectral resolution with the
isotopically pure \ybodd and \ybeven ovens, a new oven with a
natural abundance isotope distribution was built, and
Fig.~\ref{fig:excitationMixedOven} shows the atomic resonance
fluorescence spectrum measured with this oven. The spectrum was
fitted to a sum of six Voigt profiles to obtain the isotope shift of
the resonance frequencies and agreement within mutual errorbars was
found with \cite{Das2005}. In this spectrum the different isotopes
are resolved, allowing for selective ionization and hence trapping
of different isotopes of ytterbium.

The resonance \Szero - \Pone, F=3/2 corresponding to the isotope
$^{173}$YbI is separated from the \Szero - \Pone resonance in
$^{172}$YbI by only \MHz{17} \cite{Das2005}. Thus it is difficult
to spectrally resolve these two isotopes. However, if loading an ion
trap with $^{172}$YbII is desired, then this is not an obstacle,
since the odd isotope's hyperfine structure prevents this ion from
being efficiently laser cooled: If no microwave radiation is present
that avoids optical pumping into a 'dark' hyperfine level (with
respect to laser light near \nm{369} driving the \Shalf - \Phalf
transition in YbII) of the electronic ground state, then it is
unlikely that $^{173}$Yb$^+$ at a thermal velocity is sufficiently
laser cooled to be trapped. Furthermore, these ions would be visible
as dark gaps in linear ion crystals. This has not been observed
during the loading of ion crystals reported in section~\ref{sec:deterministicLoading}.

\section{One dimensional heat model of the oven}\label{sec:ovenmodel}
We have manufactured a variety of ovens with different geometries
which all have in common that ytterbium inside a metal tube with one
open end is evaporated by ohmic heating.

The latest ovens were designed to meet the special demand of a micro
structured segmented linear ion trap. To avoid coverage of the trap
chip with neutral ytterbium, the atomic beam should be restricted to
a region narrower than the electrode separation. Initial experiments
with squeezed oven nozzles gave good collimation results, confirmed
by analysing transverse Doppler-broadening profiles. But within 30
hours of operation the Doppler width broadened gradually, most
likely due to clogging or re-deposition in the nozzle. Satisfying
long term behaviour could be achieved by a laser cut slit aperture a
few millimetres apart from the oven nozzle.

\begin{figure}[htbp]
  \centering
  \epsfig{file=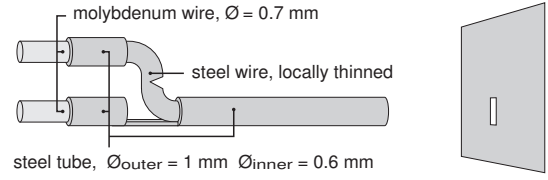, width=7cm}
   \caption{Sketch of the oven}
  \label{fig:ovenSketch}
\end{figure}

Our recent ovens are made of a steel tube and wire, which is thinned
at one point to give a high local ohmic loss. A steel tube,
partially filled with a small amount of ytterbium is mounted close
to this point and a collimated ytterbium beam emerges from the open
tube end. The feed wires are made of molybdenum.

A simple heat model for a small piece of wire states that the net flux of power per length element vanishes for a steady state temperature distribution
\begin{equation}
dP_{\rm{ohm}}  - dP_{\rm{rad}} - dP_{\rm{cond}} = 0
\label{eq:powerBalance}
\end{equation}
with the infinitesimal power contributions $dP$ due to ohmic heating, radiation, and conduction.
These powers for a small length interval $dx$ of a cylindrical wire are given as
\begin{align}
\label{eq:pohm}
dP_{\rm{ohm}} &= I^2 dR = \frac{I^2 \rho(x)}{\pi r^2} dx\\[2mm]
\label{eq:prad}
dP_{\rm{rad}} &= dA \sigma \epsilon T^4 = 2 \pi r \sigma \epsilon \left(T^4-T_0^4 \right)dx\\[2mm]
\label{eq:pcond}
dP_{\rm{cond}} &= dP_{\rm{cond}}(x-dx/2) - dP_{\rm{cond}}(x-dx/2) \nonumber\\[1mm]
&=\kappa \pi r^2 \left(T'(x+dx/2) - T'(x-dx/2)\right) \nonumber\\[1mm]
&= \kappa \pi r^2 T''(x) dx
\end{align}

In the eq. (\ref{eq:pohm})-(\ref{eq:pcond}) the heating current is
denoted as $I$, $\rho(x)$ is the electrical resistivity, $r$ is the
radius of the wire, $\sigma$ the Stefan-Boltzmann-constant,
$\epsilon$ the emissivity of the wire material and $T_0$ the ambient
temperature, and the temperature conductivity is termed as $\kappa$.
Inserting eq. (\ref{eq:pohm}-\ref{eq:pcond}) into
\eqref{eq:powerBalance} one finds that the second derivative of the
temperature has its origin in ohmic heating and radiative losses:
\begin{equation}
- \kappa \pi r^2 T'' (x) + \frac{I^2 \rho(x)}{\pi r^2} - 2 \pi r \sigma \epsilon \left(T^4-T_0^4 \right) =0
\label{eq:heatmodel}
\end{equation}
When the oven is modelled as a one dimensional  system made of the
feed through wire with a small incision as the heating element the
electric resistivity along the wire is given by
\begin{equation}
\rho(x) = \rho_0 + R \pi r^2 \delta(x)
\label{eq:resistivity}
\end{equation}
where $\rho_0$ is the resistivity of the  wire, and $R$ the
resistance of the incision at $x=0$ (treated as point like). We
integrate eq. \eqref{eq:heatmodel} numerically, which allows us to
account for the temperature dependence of electric and heat
conductivity. We take into account two boundary conditions: i) the
temperature at holders at $x=\pm \ell$ equals the ambient
temperature $T_0$, and ii) the temperature distribution has a kink
at the heating element at $x=0$ and the derivative changes by $I^2
R/\kappa \pi r^2$.

\begin{figure}[htbp]
  \centering
  \epsfig{file=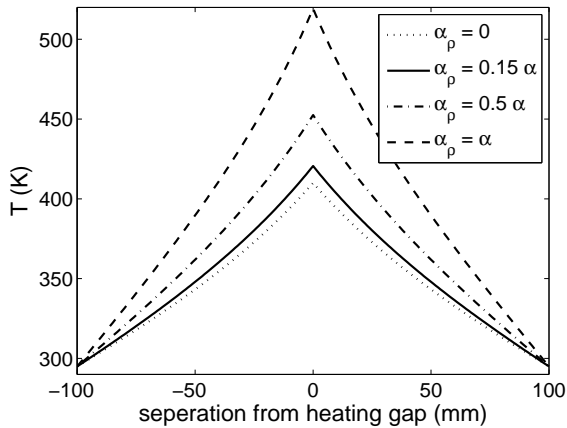, width=8cm}
   \caption{Temperature distribution along a wire with an incision at $x=0$ and held at room temperature at $x=\pm \unit[10]{cm}$}
  \label{fig:tdistribution}
\end{figure}

Figure~\ref{fig:tdistribution} shows a numerical solution for the
temperature distribution using a current of \unit[1.5]{A} and
parameters that describe our oven adequately. These parameters are
either obtained from literature (as for $\kappa$ and $\rho$) or from
our own measurements (as for $R$). In addition, we take into account
the linear temperature coefficient of the electric resistivity. The
parameters used for calculating and plotting the temperature
distribution are: $r=\unit[0.35]{mm}$, $\ell=\unit[10]{cm}$,
$R=\unit[0.06]{\Omega}$, $\kappa=\unit[138]{W/mK}$ \cite{Lide2007},
$\rho=\unit[5.57 \cdot 10^{-8} (1+4.82 \cdot 10^{-3} (T-T_0
)/K)]{m\Omega}$ \cite{Holmwood1965}. Figure~\ref{fig:tdistribution}
shows the temperature distribution for various linear temperature
dependencies of the heat conductivity. The distribution is found to
be dominantly triangular, with negligible ohmic losses in the wire
and small positive curvature for higher temperatures regions due to
radiative losses. The model gives reasonable agreement with
measurements of the oven tip temperature considering that we did not
take into account the geometric details of our oven. Best agreement
is reached when a temperature dependence of the heat conductivity is
assumed that is approximately a factor 6 lower than the coefficient
for the resistivity (as it is for iron).

\begin{figure}[htbp]
  \centering
  \epsfig{file=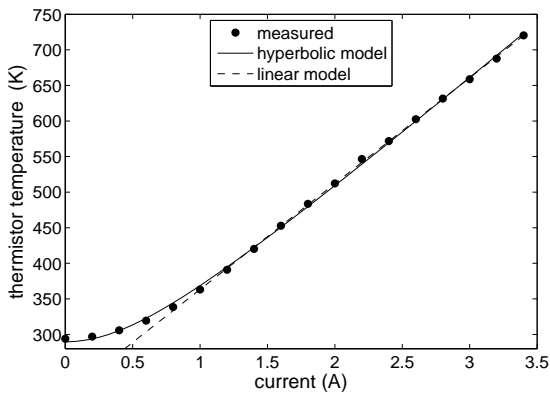, width=8cm}
   \caption{Temperature of the oven tip as a function of heating current.}
  \label{fig:thermistor}
\end{figure}

The oven tip temperature was measured close to the incision shown in
Fig.~\ref{fig:ovenSketch} using a j-type thermistor as a function of
heating current (see Fig.~\ref{fig:thermistor}) and a hyperbolic
dependence is found. This functional dependence is unexpected at
first sight: one would expect a quadratic behaviour for small
currents, as the heating powers depends quadratically on the
current, and radiative losses are negligible. In fact, such a
dependence is found for small currents. For high currents one would
expect a square root dependence of the tip temperature, arguing that
dominating radiative losses which scale as $T^4$ are balanced by a
heating power behaving as $I^2$ but such a regime is not observed
for the currents that we applied. An extended linear regime can be
explained, however, taking into account the temperature distribution
along the wire as above. The hyperbolic dependence is reproduced by
a simple analytical model assuming a triangular temperature
distribution in accordance with the results shown in
Fig.~\ref{fig:tdistribution}:
\begin{equation}
T(x) = T_{\rm{max}} - \frac{T_{\rm{max}}-T_0}{\ell} |x|\qquad |x|\leq \ell
\label{eq:ttriangular}
\end{equation}
with the ambient temperature $T_0$. The heat loss through the wire-ends at $x=\rm \ell$ equals $2\kappa
\pi r^2 T'$ and power balance requires
\begin{align}
I^2 R &= 2\kappa \pi r^2 T' + \int_{-\ell}^\ell 2 \pi r \sigma \epsilon \left( T(x)^4 -T_0^4 \right) dx     \\
&= 2\kappa \pi r^2 T' + \frac{4 \pi r \sigma \epsilon \ell}{5 (T_{\rm{max}}-T_0)}\left(T_{\rm{max}}^5 - T_0^5 \right)
\label{eq:TmaxInt}
\end{align}
After polynome division of the radiative term and considering terms up to second order in $T_{\rm{max}}$ the peak temperature is obtained as
\begin{equation}
T_{\rm{max}} \approx \left(\frac{T_0}{2} - \alpha\right) + \sqrt{\left(\frac{T_0}{2} + \alpha\right)^2 + \beta I^2}
\label{eq:tmax}
\end{equation}
with the constants $\alpha$ and $\beta$ given by
\begin{equation}
    \alpha = \frac{\kappa}{8 \ell^2 T_0^2 \epsilon \sigma} \qquad
    \beta = \frac{R}{\pi r \lambda} \alpha
    \label{eq:tcoeff}
\end{equation}
which gives the observed hyperbolic dependence. We have measured the
temperature of the oven tip with a vacuum compatible thermistor and
find good agreement with a hyperbolic model. For a current above
\unit[1]{A} (that leads to a useful atom flux as will be reported
below), a linear relation between current and temperature is found.
It should be noted that the measured quantitative asymptotic
behaviour is not predicted by the coefficients extracted from
\eqref{eq:tcoeff}, most likely due to geometric details and material
properties that are neglected or simplified in our model. However,
the important general insight remains that heat conductivity and
radiative losses yield a polynomial expression in $T_{\rm{max}}-T_0$
in eq. \eqref{eq:TmaxInt} which results in a hyperbolic dependence of
the maximum temperature as a function of heating current.

\section{Neutral Atom Densities and Atomic Flux}\label{sec:densityflux}

The temperature of ytterbium in a closed oven would result in a
vapor pressure, which, for a temperature independent latent heat, is
predicted by the Clau\-si\-us-Cla\-pey\-ron equation to take the simple form
\begin{equation}
p_{\rm{vp}}(T) = p_0 \, e^{-T_1 / T} \approx \unit[353334]{Pa} \cdot e^{-\unit[9676]{K}/T}
\label{eq:vaporPressure}
\end{equation}
where the numerical constants on the right-hand-side were obtained
by  to match measured vapor pressures in the temperature range
\unit[623]{K} to \unit[931]{K} \cite{Habermann1964}. We used this
relation despite the extended temperature range investigated here
when modelling atomic density and flux data. Assuming the ytterbium
vapor to behave as an ideal gas, the particle density can be
calculated from the vapor pressure
\begin{equation}
n = \frac{p_{\rm{vp}}(T)}{k_{\rm{B}}T} = \frac{p_0 \, e^{-T_1 / T}}{k_{\rm{B}}T}
\label{eq:density}
\end{equation}
Furthermore, for Maxwell distributed atom velocities, one can calculate the flux of atoms through an circular aperture with radius $r$
\begin{equation}
\dot{n} = \sqrt{\frac{8\pi r^4}{m k_{\rm{B}}T}}\,p_0 \, e^{-T_1 / T} = \frac{\pi r^2}{k_{\rm{B}}T} \left<v\right> \,p_0 \, e^{-T_1 / T}
\label{eq:flux}
\end{equation}
with the mean velocity of the atoms in the beam
\begin{equation}
\left<v\right> = \sqrt{\frac{8 k_{\rm{B}}T}{\pi m }}
\label{eq:beamvelocity}
\end{equation}
When calculating densities and fluxes from currents using equations
\eqref{eq:tmax} and (\ref{eq:vaporPressure} - \ref{eq:flux}), one
has to bear in mind, that the effective temperature might in general
be lower than the one obtained for the heating element, since the
atoms can thermalize with the whole, and generally cooler, oven
tube.

Atom densities and fluxes were characterized for the oven which was
optimized for use with a micro-structured segmented linear Paul
trap. The neutral atom fluorescence is imaged onto a small aperture,
which in turn is imaged onto a photomultiplier. Magnification and
aperture are chosen such that the intensity and atom density can be
assumed approximately constant over the detected area. We take into
account saturation by determining experimentally the intensity
profile of the laser and the total power. For all measurements we
assume a steady state temperature distribution which is ensured by a
10 minute warm-up.

\begin{figure}[htbp]
  \centering
  \epsfig{file=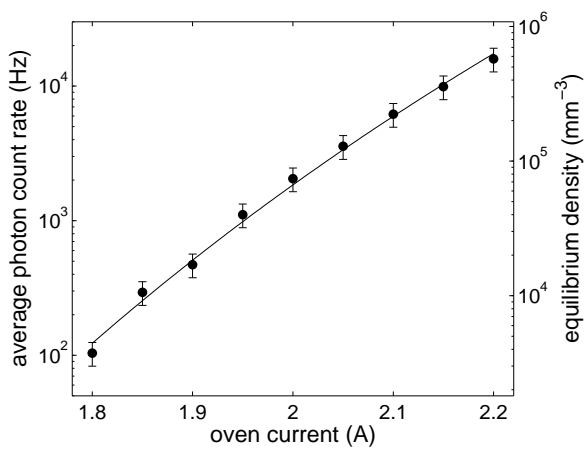, width=8cm}
   \caption{The atom density from a collimated ytterbium with natural isotope
   abundance as a function of oven heating current.
   The beam is collimated by an aperture separated \unit[3]{mm} from the oven nozzle. The density was measured
   approximately \unit[5]{mm} away from the collimating aperture.}
  \label{fig:densityDaniel}
\end{figure}

\begin{figure}[htbp]
  \centering
  \epsfig{file=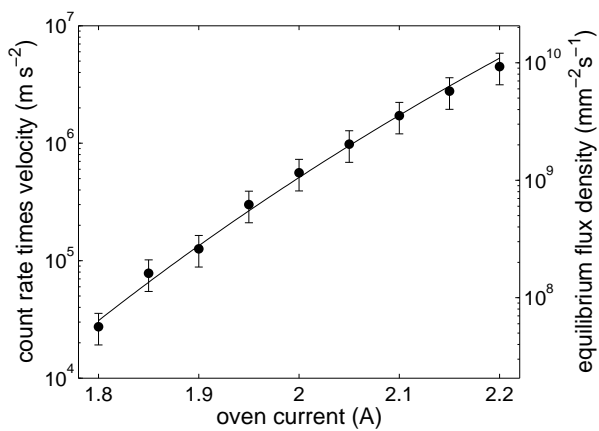, width=8cm}
   \caption{The atomic flux density from a collimated ytterbium  with natural isotope abundance as a function of oven heating current. The density data presented in Fig.~\ref{fig:densityDaniel} was combined with measurements of the line shift due to the mean beam velocity as shown in Fig.~\ref{fig:excitationPureOven1}a).}
  \label{fig:fluxDaniel}
\end{figure}

The measured current dependence (Fig.~\ref{fig:densityDaniel} and \ref{fig:fluxDaniel}) of the density
and flux is fitted by a simple model: the atomic vapour is in
thermal equilibrium with the oven. The oven has an effective
temperature, which is assumed to exhibit a linear dependence on
current as found with thermistor measurements for higher currents
(compare Fig.~\ref{fig:thermistor}). The dependence of the effective oven temperature
is not necessarily identical with the dependence of the oven tip
temperature: we allowed for a temperature offset and a different
slope; in general, the effective temperature is expected to be lower
than the one measured by the thermistor. The temperature is taken to
calculate the vapour pressure and flux. Both quantities are
multiplied by a factor, taking into account the finite atomic beam
divergence, non-equilibrium effects, imperfect characterization of
our detection efficiency and saturation.

The model describes the measurements quite well. It states that the
effective temperature of the oven (and thus the atoms' temperature)
shows the same slope with respect to the current, but is
substantially lower than the temperature of the oven tip (by
approximately \unit[50]{K}).

\section{Two-color-ionization of YbI}
\label{sec:twoColor} In this section we report on resonance enhanced
isotope selective ionization of YbI with subsequent loading of the
ions into a Paul trap (ring trap or linear trap). The possibility to
selectively excite a certain isotope out of a mixture of natural
abundance material is an advantage of photoionization. This
selectivity is achieved by the first excitation step from the ground
state of YbI into the \Pone state as reported in the previous
section. The second excitation step from the \Pone state of YbI into
the ionization continuum is achieved by means of laser light near
\nm{369} (Fig.~\ref{fig:ionizationScheme}). The dependence of trap
loading efficiency on laser intensities, laser frequencies, and the
relative flux of neutral atoms is studied.

\subsection{Experimental procedure}
\label{sec:experiment} A part of the experimental setup is described
already in the preceding section that deals with resonant excitation
of YbI. Now, taking advantage of this isotope selective excitation
of the \Pone state, we are concerned with photoionization of YbI and
trapping of YbII. Ions are trapped in either a miniature ring Paul
trap or a linear Paul trap. The ring trap electrodes are made from
molybdenum wire with a diameter of \mm{0.3}. The ring electrode has
an inner diameter of \mm{2.0} and the endcap electrodes are \mm{1.4}
apart. An rf field at \MHz{9.5} with peak amplitude of about
\unit[700]{V} is applied between ring and endcap electrodes. The
linear trap is made of four Molybdenum rod electrodes \mm{0.5} in
diameter, held by two Macor plates spaced \mm{6} apart. The radial
distance from the axis to the surface of the rods is \mm{0.75}. Two
Molybdenum end-caps with a diameter of \mm{0.4} serve for axial
confinement, and are spaced \mm{4.1} apart. An rf field with
frequency between 10 and \MHz{20} and amplitude of up to
\unit[2]{kV} is applied to the rod electrodes for radial
confinement, and a low voltage of several volts only is applied for
axial confinement. Typical secular frequencies are \kHz{30-60}  in
the axial direction and \kHz{600-800} radially.

Light near \nm{369} driving the \Shalf - \Phalf transition in YbII is
supplied by a frequency doubled commercial Ti:Sa laser (Coherent
MBR110).  This will be termed "the cooling laser" in the remainder
of this article. A diode laser delivers light near \nm{935} and drives
the D$_{3/2}$ - [3/2]$_{1/2}$ transition in YbII to avoid optical
pumping into the metastable D$_{3/2}$ state \cite{Bell1991}.

An electric current resistively heats the oven, and after some time,
a stationary temperature is obtained. Therefore each oven was heated
for at least \unit[10]{min} prior to the actual experiment, unless
stated otherwise. The oven temperature determines the vapour
pressure and atom density inside the oven and the atomic flux
density in the particle beam that emerges from the open end of the
oven.

In order to determine quantitatively the atomic number density in
the trapping region of the ion trap one would need knowledge of i)
the spatial profile of the atomic beam and the laser beam in the
region where they cross, ii) the overall detection efficiency of the
system measuring the resonance fluorescence in this crossing region
as function of the position of the fluorescing atoms (the depth of
focus for the optics collecting the fluorescence is rather small due
to a large numerical aperture), and iii) the transition line
strength of the \Szero - \Pone transition. Thus, deducing the number
density from the observed resonance fluorescence spectra is in
principle feasible, albeit with a large systematic error. If the
number density were known, the investigation of ion loading rates as
a function of oven current reported below could be presented as a
function of neutral atom density instead. The ion loading rates
depend, in addition on i) the ionization cross section for the
different ionization schemes reported (one-color, two-color,
electron impact), ii) the quasi static electric field experienced by
the atoms in the trap region that varies spatially and in time in a
different manner for each trap, and iii) the rate at which different
YbII ions are laser cooled which in turn is a function of the
spatial profiles and overlap of the laser beams near \nm{369} and
\nm{935} used for laser cooling YbII and their frequency, intensity,
and polarization (in addition, for odd isotopes also of the
parameters characterizing the microwave radiation necessary to avoid
optical pumping). We do not attempt here to give a full quantitative
description that might be plagued by a large uncertainty while not
altering the conclusions drawn from the measurements reported below.

The wavelength of the laser near \nm{398.9} was set to a value
corresponding to the \Szero - \Pone resonance of the desired isotope
to be loaded into the trap, and the peak laser intensity (Gaussian
beam profile) in the ionization region was \unit[3]{W/cm$^2$}, if
not explicitly stated otherwise. For brevity we will label this
laser "excitation laser" in what follows.

When both the laser near \nm{369} and near \nm{398.9} are directed
into the trap region the atomic resonance fluorescence signal is
added on top of the ionic fluorescence. Experimentally the signals
due to YbI and YbII are discernible by blocking temporarily the
repump laser at \nm{935} which leads to optical pumping into the
metastable D$_{3/2}$ state of YbII and thus the ionic fluorescence
is interrupted. Typically, the fluorescence signal collected from
neutral atoms amounts to - depending on oven current - several tens
of kHz.

\subsection{Two-color ionization}

\begin{figure}[htbp]
  \centering
  \epsfig{file=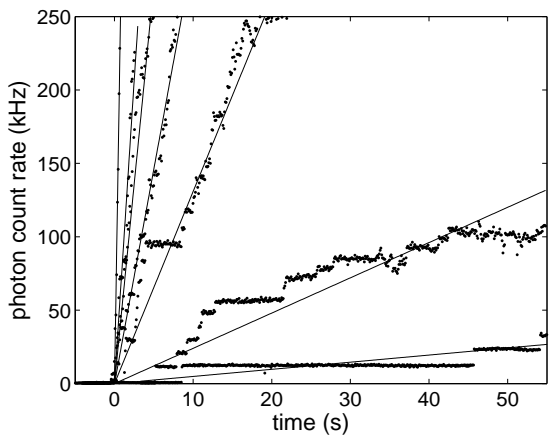, width=8cm}
  \caption{Two-color  photoionization of $^{172}$YbI as a
    function of time for different electric currents heating the atom
    oven. The fluorescence signal is proportional to the number of
    trapped YbII ions for a low number of ions, i.e., short loading times
    (here, a single ion produces a photon count rate of $\approx$ \kHz{15}).
    In this region the signals are fitted by a linear function to extract
    loading rates as a function of oven current. Steps in the photon count rate
   for small heating currents indicate an increase by one of the number of
   trapped ions.
   The oven currents were (in order of ascending signal slope):
   \unit[1.8]{A}, \unit[1.9]{A}, \unit[2.0]{A}, \unit[2.18]{A},
   \unit[2.3]{A}, \unit[2.47]{A}, \unit[2.57]{A}.}
  \label{fig:twoColorVsCurrent}
\end{figure}

Fig.~\ref{fig:twoColorVsCurrent} shows combined atomic (YbI) and
ionic (YbII) fluorescence signals in dependence on the ionization
time and the oven current for two-color-ionization. The procedure to
record the time series in Fig.~\ref{fig:twoColorVsCurrent} is as
follows: The ion trap is emptied by turning off the rf trapping
field and all lasers except the excitation laser at \nm{398.9} are
turned on. After \unit[3]{s} the excitation laser is unblocked,
leading to fast loading of the trap and accordingly to an increase
in the ionic fluorescence signal. With a fluorescence rate of
\kHz{15} for a single ion here, on the order of a 100 ions are
trapped in the first 15 seconds for the highest oven current at
\unit[2.57]{A}, indicating the high efficiency of this process. This
oven current represents the usual value used with electron impact
ionization.

The resonance fluorescence signal begins to saturate, after a time
depending on oven current, generally for ion numbers greater than
about 20, owing both to the limited trapping capacity and the
reduced efficiency of the optics for imaging a large ion ''cloud''
(the optical set up used here is optimized for imaging a single ion
located in the center of the trap). A limited capacity for trapping
can be attributed to the laser beam waist being too small to
illuminate large clouds. In addition,   large hot clouds exhibit a large number
of micromotion sidebands, such that the scattering of resonance
fluorescence induced by a laser with narrow emission bandwidth is
reduced dramatically.

\begin{figure}[htbp]
  \centering
  \epsfig{file=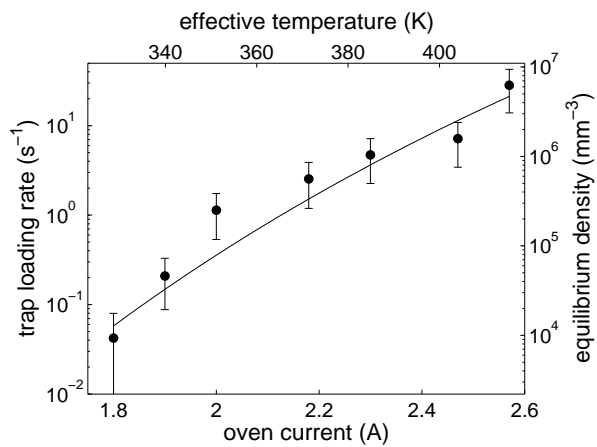, width=8cm}
   \caption{Loading rate extracted from the slope of fluorescence
   increase in the linear regime as a function of oven current
   (compare Fig.~\ref{fig:twoColorVsCurrent}).
   A thermodynamic vapor density is fitted to the data assuming
 a linear dependence between current and effective temperature
 as observed from probe measurements (see
section~\ref{sec:densityflux}). The fitted dependency is comparable
to the one measured in section~\ref{sec:densityflux}, even though
the oven was constructed with a slightly different design.}
  \label{fig:loadingRate}
\end{figure}

The oven current is varied between \unit[1.80]{A} and
\unit[2.57]{A}. For those currents we obtained feasible loading
rates and low vacuum contamination. As expected, the rate of loading
ions increases with the atom flux density. Depending on the
probability for ionizing a neutral atom passing by, two dependencies
are conceivable: for almost unit probability of ionization, the
loading rate will be proportional to the atom flux, whereas for low
probability of ionization, the loading rate can be expected to
behave proportional to the atom density. A comparison of loading
rates on the order of a few per second and much higher measured
fluxes shows that the latter is the appropriate model, and
Fig.~\ref{fig:loadingRate} shows the measured loading rate together
with a fit that assumes proportionality to the atom density.

\begin{figure}[htbp]
  \centering
  \epsfig{file=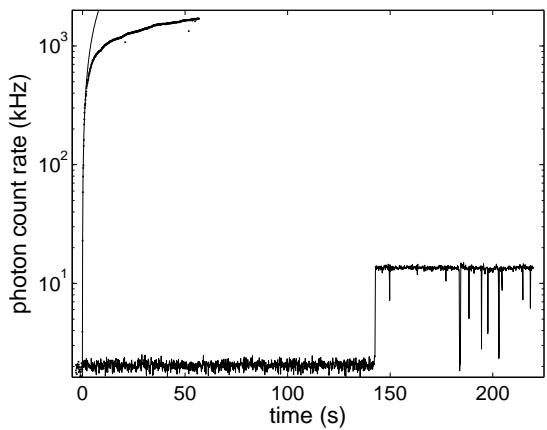, width=8cm}
  \caption{Comparison of efficiency for electron impact and two-color-ionization of $^{172}$YbI
  (oven current set to \unit[2.5]{A}) \cite{Balzer2006}. The upper trace (dots)
shows the rise in
  fluorescence when two color photoionization is used and a fit (solid line) to the linear slope of
  the signal as in Fig.~\ref{fig:twoColorVsCurrent}
  (note that the linear slope is not displayed as a straight line due to the logarithmic depiction).
  The step in the lower time trace (electron impact ionization)
  signifies the trapping of a single ion.}
  \label{fig:comparisonElectronVsTwoColor}
\end{figure}

In Fig.~\ref{fig:comparisonElectronVsTwoColor} the loading rate using
two-color-photoionization is compared with electron impact
ionization. The time series indicating the fluorescence signal when
using electron impact ionization shows a sharp increase at about
142\,s that signifies the trapping of a {\em single} \ybeven ion. In
contrast, the ionic signal rises quickly when using photoionization
indicating trapping of more than 100 ions during the first seconds
after turning on the excitation laser. The two-color-ionization
process results in an estimated increase in trap loading rate of
about 3000 in comparison with electron impact ionization. For this
measurement the oven current was set to \unit[2.5]{A}.

In the two paragraphs to follow we examine the influence of the
frequency of both the laser light field near \nm{369} and the one near
\nm{399} on the efficiency of the ionization process. The ionization
model depicted in Fig.~\ref{fig:ionizationScheme} is corroborated by
these measurements.

\begin{figure}[htbp]
  \centering
  \epsfig{file=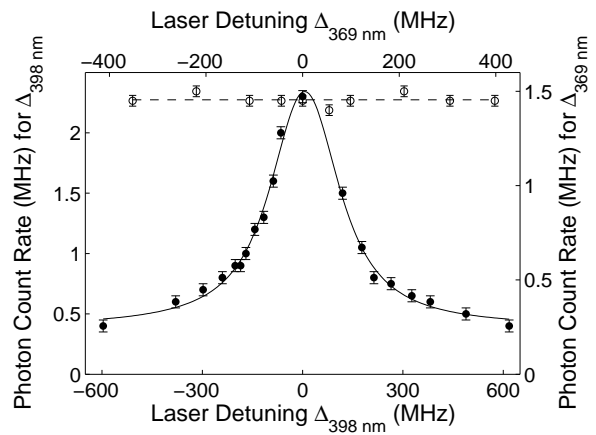, width=8cm}
  \caption{
  Dependence of ionic resonance fluorescence signal - and hence the
   loading rate - on detuning of both lasers for two-colour-photoionization during the loading time.
   The fluorescence signal shows a strong dependence on the detuning of the
   \nm{398} laser (filled dots), since this excitation is a resonant
   process. The solid line indicates a fit using a
   Voigt profile.
   The non-resonant excitation with the \nm{369} laser above ionization
   threshold shows no dependence upon variation of wavelength
   (open circles) and can be fitted by a constant (dashed line).
   The ionization time was \unit[30]{s} at \unit[2.59]{A}
   oven current.}
  \label{fig:selectivity369}
\end{figure}

Fig.~\ref{fig:selectivity369} shows the resonance fluorescence
signal of trapped \ybeven ions under variation of the frequency of
the cooling laser close to \nm{369} over a range of \MHz{800} during
the loading phase. The scan was taken with the oven current set to
\unit[2.59]{A} and \unit[30]{s} ionization time. During the
observation stage of the trapped ions, the laser at \nm{369} was set
close to resonance. The observed fluorescence signal is consistent
with a structureless flat dependence of the ionization rate on the
detuning of the cooling laser in the frequency range depicted here.
This is in accord with the ionization model that the atom is excited
from the \Pone state into the ionization continuum
(Fig.~\ref{fig:ionizationScheme}). The energy of the photon at
\nm{369} is larger than the energy needed to reach the continuum
(equivalent to a wavelength of \nm{394}), thus no dependence on
small frequency variations is expected.

On the other hand, based on the excitation spectra of neutral YbI
the detuning of the excitation laser at \nm{398.9} is expected to
have a distinct influence on the ionization efficiency for a given
isotope. The second data set in Fig.~\ref{fig:selectivity369}
confirms this assumption. There, the excitation laser is scanned
over a range of 1.2 GHz and the resulting {\em ionic} fluorescence
is recorded. This graph may also be compared to
Fig.~\ref{fig:excitationPureOven1}a, since both data sets were taken
using the \ybodd isotope and an angle of 45 deg between the atomic
and laser beam, respectively (thus the line profile is dominated by
Doppler-broadening). The oven current was set to \unit[2.82]{A} and
the time during which the excitation laser was turned on was
\unit[60]{s}.

\subsection{Nearly deterministic loading of an ion trap}
\label{sec:deterministicLoading} The last paragraph of this section
deals with an experimentally important feature of photoionization of
YbI for ion trap experiments: It is desirable to (nearly)
deterministically load a desired number of ions into the trap. In
addition, the flux of neutral atoms possibly contaminating trap
structures should be significantly reduced. So far, the oven current
was set to a certain value and kept there to allow for comparison of
different loading rates. Now, when trying to quickly load ions, all
laser frequencies and intensities are set first before the oven
current is turned on. The wavelength of the excitation laser near
\nm{399} is set with an accuracy better than \MHz{80} using a wave
meter, which is sufficient to find the atomic resonance without
having to rely on the atomic beam. Thus, already the first atoms
emitted from the oven have a chance of being ionized by efficient
two-color-ionization and are immediately laser cooled by the same
laser light that ionizes them. It turns out that the loading rate
using this procedure is such that any desired (low) number of ions
is easily selected simply by blocking the ionization laser at the
right time.

\begin{figure}[htbp]
  \centering
  \includegraphics[width=8cm]{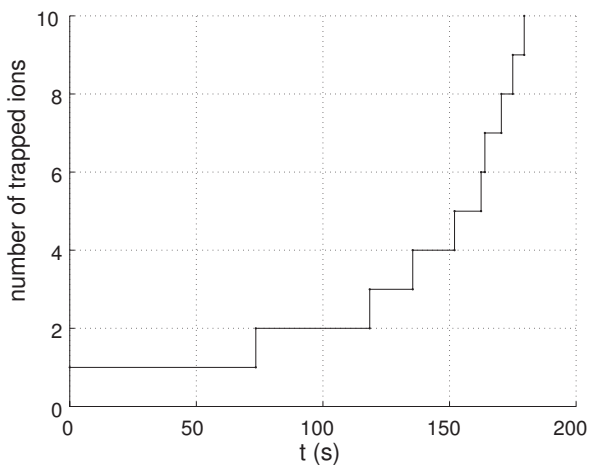}
  \caption{Deterministic loading of a desired number of ions by two-color-photoionization \cite{Balzer2006}.
  The loading rate increases with time, since
  the oven emitting YbI atoms heats up during the measurement}
  \label{fig:deterministicLoading}
\end{figure}
Fig.~\ref{fig:deterministicLoading} shows the number of trapped ions
plotted against time. This experiment was performed with a linear
Paul trap and the ionic fluorescence was detected by an
image-intensified CCD-camera. The time before the first ion is
detected varies between \unit[60 - 180]{s} after the oven-current is
turned on, depending on laser parameters. Then, while the oven is
being heated the loading rate increases. This graph shows that
nearly deterministic loading of a desired number of ions is possible
using this procedure.

\section{Single-color-ionization of YbI}
\label{sec:oneColor} The energy carried by two photons at \nm{398.9}
is not sufficient to bridge the energy gap between the ground state
of YbI and the ionization continuum. Therefore, it came initially as
a surprise that YbII ions could be produced using this laser light
alone.

For the experiments presented in this section all lasers are blocked
first for a prescribed time while the laser at \nm{399} is on. Then
the laser exciting the atomic resonance is blocked while lasers near
\nm{935} and \nm{369} are turned on for cooling and detecting the ion.

Comparing the two methods shows that for the experimental parameters
used here, the two-color-ionization is almost two orders of
magnitude more effective than the one-color-ionization. In the
latter case, though, the ions are cooled only after the ionization
process is finished when the cooling lasers are unblocked. For the
two-color scheme, on the other hand, the ions are cooled as soon as
they are ionized.

As an explanation of why ions are produced by \nm{399} light alone,
3-photon-ionization would be conceivable, with the first photon
exciting the atomic \Szero - \Pone resonance as before, and two more
photons ionizing the atom. This "second" two-photon process could
again be resonance enhanced, if the \nm{399} light excites a Rydberg
state in YbI. If the absorption of the last two photons is {\em not}
resonance enhanced, then a rough estimate shows that at a light
intensity of a few Watts per cm$^2$ this three-photon-ionization
rate (proceeding resonantly with the first photon through the \Pone
state and then with two more photons {\em non} resonantly into the
continuum) should be about 12 orders of magnitude lower than the
two-photon rate (i.e., proceeding resonantly with the first photon
through the \Pone state and then with {\em one} more photon into the
continuum) \cite{Chin1984}. Since such a low ionization rate is not
consistent with our experimental observation, we conclude that a
three-photon process that is not a second time resonance enhanced
cannot be responsible for the observed production of YbII using
light near \nm{399}.

Is there a Rydberg state that could be resonantly excited starting
from the \Pone state with the same light that excites the \Szero -
\Pone transition, and thus lead to a three photon process that is
resonantly enhanced a second time? From this Rydberg state, a third
photon at the same wavelength would then ionize the atom. Starting
from the \Pone level the dipole-selection rules allow transitions
into either S- or D-Rydberg states. Checking the energy levels of
the S- and D-Rydberg series shows that no level exists at
\unit[50136.444]{cm$^{-1}$}, the energy corresponding to two photons
near \nm{398.9} \cite{Camus1980,Xu1994}. The nearest unshifted
singlet states are the 6s23s$^1$S$_0$ at \unit[50130.98]{cm$^{-1}$}
and the 6s22d$^1$D$_2$ at \unit[50148.59]{cm$^{-1}$}, with an energy
difference to the value above of \unit[5.46]{cm$^{-1}$} and
\unit[12.15]{cm$^{-1}$}, respectively. We therefore exclude resonant
excitation of an {\em unshifted} Rydberg state as a likely candidate
for the observed photoionization by a one-color light-field.

When considering possible ionization pathways we have neglected so
far the influence of the rf trapping field on the ionization
process. Given the trap dimension of the ring trap and the applied
rf voltage the peak electric field is estimated as \unit[$7\times 10^5$]{
V/m}. With the Stark shift taking on a value of about \unit[$10^{-6}$]{
Hz/(V/m)$^2$} for Hydrogen with $n=2$ and being proportional to $n^7$
(with principal quantum number $n$) \cite{Demtroder2003} it is
expected that the Stark shift of the \Pone state is negligible here
while Rydberg states with $n\approx 22$ are considerably shifted in
energy. In fact, the trapping potential lowers the ionization
threshold such that a second photon at \nm{398.9} directly reaches
the continuum.

In a simplified picture a Hydrogen atom exposed to a
constant external electric field in direction $x$ shows an asymmetric binding potential along this direction as
\begin{equation}
V = -\frac{e^2}{4\pi\epsilon_0} \frac 1x - e E x,
\end{equation}
with elementary charge $e$, electron nucleus separation $x$, and the magnitude of the external
electric field $E$. The potential maximum in the direction of the electric field is lowered with respect to the unperturbed atom which corresponds to a lowering of the ionization
threshold by
\begin{equation}
  \label{eq:schottky}
  \Delta U = \sqrt{\frac{e^3 E}{\pi \epsilon_0}}.
\end{equation}

This is reminiscent of the Schottky effect \cite{Schottky1914} which
describes the lowering of the work function of solid state materials
in the presence of an electric field.

The difference of the ionization energy $U =$ \ka{50443} and the energy of two photons $2 E_{398} = $ \ka{50136.4} requires a lowering of the ionization energy by at least $\Delta U = $ \ka{306.6}. Using eq. \eqref{eq:schottky}, this corresponds to a required field strength exceeding $|\vec{E}_{\rm{ph}}| =$~\unit[$2.5\cdot 10^5$]{V/m}.

On the other hand the peak potential in a linear ion trap along the $z$ direction is given by
\begin{equation}
  \phi_{\rm{peak}} = U_0 \frac{x^2 - y^2}{r_0^2}.
\end{equation}
with the peak voltage $U_0$ and the electrode separation $2 r_0$ yielding a peak electric field of
\begin{equation}
  E_{\rm{peak}} = -\left|\vec{\nabla} \phi\right| = \frac{2 U_0}{r_0^2}\sqrt{x^2+y^2} = 2 U_0 \frac{r}{r_0^2}.
\end{equation}
Thus the radius, at which the required electric field strength for one-color-photoionization is exceeded amounts to
\begin{equation}
  r = \frac{|\vec{E}_{\rm{ph}|} r_0^2}{2 U_0},
\end{equation}
which gives for our trap ($U_0 = \unit[1000]{V}$ and $2\, r_0 = \mm{1.5}$) a separation from the trap center of
$r=$~\mum{140}. This gives \unit[.75]{eV} of initial quasi potential energy which has to be laser cooled.

A full quantitative treatment of the electric field enhanced
photoionization has to account, in particular, for the density and
velocity distributions of the YbI atoms moving through the
oscillatory trapping field, and the exact spatial dependence of the
trapping field.

The significantly lower trap loading rate when using \nm{399} light
alone for photoionization, as compared to the two-color process, is
explained when considering i) that only a fraction of the atoms
interacting with \nm{399} light has their ionization threshold
sufficiently lowered to be ionized, and ii) the fact that during the
short time interval between turning off the laser near \nm{399} and
turning on the lasers necessary for cooling YbII a fraction of the
ions will have already left the trapping region.

\begin{figure}[htbp]
  \centering
  \epsfig{file=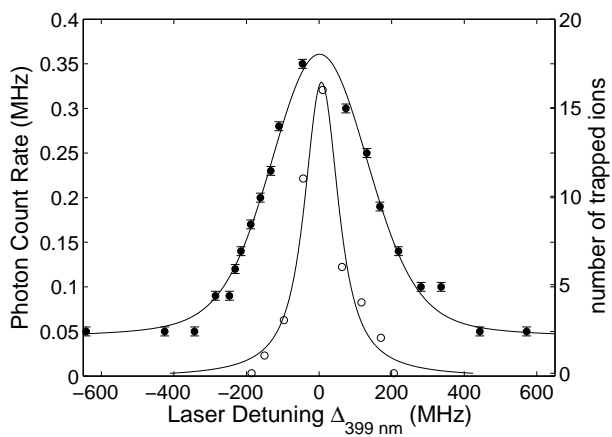, width=8cm}
  \caption{Ionization efficiency of one-colour photoionization of
  \ybeven as a function of detuning of the \nm{399} light from the
  \Szero - \Pone resonance. The filled circles show the fluorescence
  signal as a function of laser detuning, similar to Fig.~\ref{fig:selectivity369}. The ionization time was \unit[30]{s} at an oven current of \unit[2.59]{A} and laser and atomic beam enclose an angle of 45 degrees. Fitting a Voigt profile yields a FWHM of \unit[320]{MHz}. \newline
The open circles display the number of trapped ions after a loading
time interval of \unit[30]{s} at \unit[2.79]{A} oven current Here
laser and atomic beam are perpendicular yielding low Dopplerbroadening. Fitting a Voigt profile yields a FWHM of \unit[110]{MHz}
which is consistent with fluorescence profiles from neutral
\ybeven.}
  \label{fig:selectivity398}
\end{figure}

Fig.~\ref{fig:selectivity398} shows the dependence of the loading
process as a function of the detuning of the light near \nm{399}
from the \Szero - \Pone resonance in $^{172}$YbI. One dataset
(filled circles) was taken with an oven current of \unit[2.59]{A}
and the ionization time of \unit[30]{s}. Due to an angle of 45
degrees between laser and atomic beam, the line width (FWHM
\MHz{320}) shows substantial Doppler-broadening (when comparing this
with Fig.~\ref{fig:excitationPureOven1}a) it should be noted that
(Fig.~\ref{fig:excitationPureOven1}a) shows a resonance fluorescence
spectrum of $^{171}$Yb, while the data in
Fig.~\ref{fig:selectivity398} shows the ionization of \ybeven as a
function of laser detuning). The second scan was taken with the
laser beam perpendicular to the atom beam, an oven current of
\unit[2.79]{A} and ionization time of \unit[30]{s}. The line width
displayed here (FWHM \MHz{110}) is consistent with the width of the
atomic resonance line recorded as shown in
Fig.~\ref{fig:excitationPureOven1}c) with $^{172}$YbI. Thus, the
ionization rate as a function of frequency of the laser light field
near \nm{399} is consistent with the absorption of the first photon
(\Szero - \Pone transition) being a resonant process and the second
photon absorption ending in the ionization continuum.

\section{Conclusion}
\label{sec:conclusion} We report on two photoionization processes
for neutral ytterbium that allow for loading of Yb$^+$-ions into a
Paul-trap. Both methods are implemented using just one additional
diode laser near \nm{399}.  Laser diodes exist for this wavelength,
and are close to the DVD-standard at \nm{405}. Since the saturation
intensity of the \Szero - \Pone transition in neutral Yb is rather
small, a diode delivering of the order 10 mW is sufficient for
efficient ionization. Thus, a usual diode laser system in external
cavity setup characterized by a small emission linewidth and easy to
tune suffices to implement efficient photoionization of Yb.

The second step in the photoionization process (after excitation of
the \Szero - \Pone resonance) is achieved by absorption of a photon
near \nm{369}, the same light field used for cooling and detection of
YbII ions. Therefore, it is ensured that the ionization process
takes place in a region where the ion is immediately laser cooled
and thus trapped and detected.

In addition, it is demonstrated that in the presence of the rf field
used for trapping ions in a Paul trap, the light field near \nm{399}
alone suffices for a second excitation step into the ionization
continuum shifted by the quasi static electric field applied to the
trap electrodes.

The isotope-shift of the \Szero - \Pone transition in neutral YbI is
resolved in excitation spectra when reducing Doppler-broadening by
crossing the atom and the laser beam at an angle near 90$^\circ$.
Thus isotope selective ionization is possible. The efficiency of
two-color-photoionization is found to be larger by a factor of 3000 in
comparison with electron impact ionization for the experimental
set-up used in these experiments. We showed that (nearly)
deterministic loading of a desired number of YbII ions into a linear
Paul trap is feasible.

We thank Andres Varon for careful reading of the manuscript.
Financial support by the Deutsche Forschungsgemeinschaft, the
European Union (Integrated Project QAP, STREP Microtrap), and by
secunet AG is gratefully acknowledged. \\

\bibliographystyle{revtex}
\bibliography{lit_Photoionisation}

\end{document}